\documentclass[aps,prb,twocolumn,groupedaddress,showpacs]{revtex4-1}
\usepackage{amsmath,amsthm,amssymb}
\usepackage[dvipdfmx]{graphicx}

\begin{document}

\newcommand{\bi}[1]{\ensuremath{\boldsymbol{#1}}} 

\title{Understanding Muon Diffusion in Perovskite Oxides below Room
Temperature Based on Harmonic Transition State Theory}

\author{T.~U.~Ito}
\affiliation{Advanced Science Research Center, Japan Atomic Energy
 Agency, Tokai, Ibaraki 319-1195, Japan}
\author{W.~Higemoto}
\affiliation{Advanced Science Research Center, Japan Atomic Energy
Agency, Tokai, Ibaraki 319-1195, Japan}
\affiliation{Department of Physics, Tokyo Institute of Technology,
Meguro, Tokyo 152-8551, Japan}
\author{K.~Shimomura} 
\affiliation{Institute of Materials Structure Science, High Energy
Accelerator Research Organization (KEK), Tsukuba,
Ibaraki 305-0801, Japan}

\date{\today}

\begin{abstract}
In positive muon spin rotation and relaxation ($\mu^+$SR) spectroscopy,
 positive muons ($\mu^+$) implanted into solid oxides are conventionally
 treated as immobile spin-probes at interstitial sites below room
 temperature. This is because each $\mu^+$ is thought to be tightly
 bound to an oxygen atom in the host lattice to form a muonic analogue
 of the hydroxy group. On the basis of this concept, anomalies in $\mu^+$SR
 spectra observed in oxides have been attributed in most cases to the
 intrinsic properties of host materials.
 On the other hand, global $\mu^+$ diffusion with an activation
 energy of $\sim$0.1~eV has been reported in some
 chemically-substituted perovskite oxides at cryogenic temperatures,
 although the reason for the small activation energy despite the formation of
 the strong O$\mu$ bond has not yet been quantitatively understood.
 In this study, we investigated interstitial $\mu^+$
 diffusion in the perovskite oxide lattice using KTaO$_3$ cubic
 perovskite as a model system. We used the $\mu^+$SR method and density
 functional theory calculations along with the harmonic
 transition state theory to study this phenomenon both
 experimentally and theoretically.
 Experimental activation energies for global $\mu^+$ diffusion
 obtained below room temperature were less than a quarter of the calculated
 classical potential barrier height for a bottleneck $\mu^+$
  transfer path.
 The reduction in the effective barrier height could be 
 explained by the harmonic transition state theory with a zero-point
 energy correction; a significant difference in zero-point energies for
 $\mu^+$ at the positions in the O$\mu$ bonding equilibrium state and a
 bond-breaking transition state was the primary cause of the reduction.
 This suggests that the assumption of immobile $\mu^+$ in solid oxides
 is not always satisfied since such a significant decrease in diffusion
 barrier height can also occur in other oxides.

\end{abstract}

%\pacs{61.72.-y, 66.30.jp, 76.75.+i}

\maketitle

%%%%%%%%%%%%%%%%%%%%%%%%%%%%%%%%%%%%
%%%%%%%%%%%%INTRODUCTION%%%%%%%%%%%%
%%%%%%%%%%%%%%%%%%%%%%%%%%%%%%%%%%%%
\section{INTRODUCTION}
The positive muon spin rotation and relaxation ($\mu^+$SR) method has
been widely used to investigate the microscopic properties of
solids.\cite{amato97,blundell04,cox09,yaounc10,higemoto16,ito20}
This technique involves implanting spin-polarized positive muons
($\mu^+$) into lattice interstices as microscopic magnetic probes to
observe local magnetic fields arising from surrounding electrons and
nuclei. This allows for the study of the electronic state of materials
from a microscopic point of view, in a manner similar to the nuclear
magnetic resonance technique.
In most cases, $\mu^+$SR is used as a silent probe; the implanted
$\mu^+$ is assumed to stay at fixed sites and not to
alter the electronic properties of the host material. However, in
reality, the implanted $\mu^+$ may change local electronic structures
and atomic arrangements to some extent due to its positive
charge.\cite{feyerherm95,tashma97,gygax03,ito09,foronda15}
In addition, $\mu^+$ is relatively mobile in crystalline lattices
because of its light mass, which is approximately one ninth that of
proton. The diffusion of $\mu^+$ can cause drastic changes in $\mu^+$SR
spectra, such as the narrowing of lines\cite{storchak98} and the averaging of
spin precession frequencies,\cite{alexandrowicz99,ito10} even when host
materials do not exhibit any anomalous behavior.
Therefore, a deep understanding of both $\mu^+$ diffusion and the
perturbation induced by $\mu^+$ on its surroundings is always
necessary for researchers using $\mu^+$SR spectroscopy, regardless of
whether they are interested in such phenomena or not.

The diffusion of $\mu^+$ has been extensively studied in elemental metals and 
its quantum nature at cryogenic temperatures has been unambiguously 
established.\cite{storchak98}
For instance, anomalous temperature dependencies of an interstitial $\mu^+$
hopping rate in copper were successfully described by a microscopic
quantum tunneling model that takes into account interactions between $\mu^+$
and the lattice, and conduction
electrons.\cite{kadono89,luke91,kondo84-1,kondo84-2,yamada84}
In contrast, $\mu^+$ implanted into solid oxides is usually regarded as
 immobile below room temperature because it is thought to be tightly
 bound to an oxygen atom in the host lattice to form a muonic analogue
 of the hydroxy group.
 Many researchers have accepted this assumption and applied $\mu^+$SR
spectroscopy to investigations of magnetism and ion
diffusion in oxides, where anomalies appearing in
$\mu^+$SR spectra have simply been interpreted as manifestations of
the material's nature.\cite{hord10,koda19,sugiyama13}
On the other hand, $local$ $\mu^+$ hopping has undoubtedly been observed
below room temperature in transition-metal oxide antiferromagnets with
the corundum structure.\cite{dehn20,dehn21} 
Moreover, $global$ $\mu^+$ diffusion with an activation energy of
$\sim$0.1~eV has been reported in some chemically-substituted perovskite
oxides, such as Sc-doped SrZrO$_3$\cite{hempelmann98} and
BaTiO$_{3-x}$H$_x$ oxyhydride\cite{ito17}, at cryogenic temperatures.
While the quantum effect associated with the light mass is inferred to be important in $\mu^+$ diffusion, the reason for the small activation energy despite 
the formation of the strong O$\mu$ bond has not yet been quantitatively
understood.

In this article, we report $\mu^+$SR and density functional theory (DFT)
calculations studies on $\mu^+$ diffusion in insulating KTaO$_3$ with 
a cubic perovskite structure to gain a comprehensive understanding of
the $\mu^+$ diffusion in the perovskite oxide lattice.
KTaO$_3$ is an ideal system for investigating the $\mu^+$ diffusion
since it has dense nuclear spins and preserves the cubic perovskite
structure even at cryogenic temperatures. These features facilitate 
experimental detection of the $\mu^+$ diffusion and its modeling in DFT
calculations.
DFT has recently been used extensively to provide valuable information
for identifying $\mu^+$ stopping sites and for estimating their stabilities
and charge states as hydrogen-like defects in crystalline lattices from a
theoretical
viewpoint.\cite{foronda15,koda19,moller13,moller13-2,vieira16,dehn20,dehn21,ito22} 
However, the application of DFT to support $\mu^+$SR
data analysis has mostly been limited to a classical level, despite the
importance of quantum effects arising from the light mass of $\mu^+$.
A recent DFT study by Onuorah {\it et al.} highlights the importance of
zero-point energies (ZPEs) in finding stable $\mu^+$ sites in
metals,\cite{onuorah19} but their impact on $\mu^+$ diffusion has
not been discussed in detail.
In this study, we handle ZPEs for $\mu^+$ in KTaO$_3$ within the
frameworks of double Born-Oppenheimer and harmonic
approximations,\cite{onuorah19} and show how the effective activation
barrier for interstitial $\mu^+$ diffusion is lowered due to the
difference in ZPEs for $\mu^+$ at positions in the O$\mu$ bonding
equilibrium state and a bond-breaking transition state based on the
harmonic transition state theory.\cite{sholl09,sholl14}

This paper is organized as follows. After a brief description of the
experiments in Sec.~\ref{secExpDetail}, we report our experimental
results in Sec.~\ref{secExpResult} together with the results of data
analysis using the dynamical Gaussian Kubo-Toyabe theory and a two-state
model. In Sec.~\ref{secComp} and \ref{secCompRes}, we describe
computational details and results of DFT calculations performed for
providing a semi-quantitative description of long-range interstitial
$\mu^+$ diffusion in KTaO$_3$ below room temperature.
In Sec.~\ref{secDisc}, we discuss the implications of these results for
interstitial $\mu^+$ diffusion in oxides with a specific focus on the
impact of ZPEs on the effective barrier height for $\mu^+$ hopping.
Finally, we summarize our conclusions in Sec.~\ref{secConc}.

\section{EXPERIMENTAL DETAILS}\label{secExpDetail}
High-purity (4N) KTaO$_3$ single crystals of
10$\times$10$\times$0.5~mm$^3$ were obtained from MTI Corporation, USA.
The single crystals were grown by the top-seed flux method and cut along
the cubic (001) plane.
Time-differential $\mu^+$SR experiments were performed at Japan Proton
Accelerator Research Complex (J-PARC), Japan, and Paul Scherrer Institut (PSI),
 Switzerland, using a spin-polarized surface muon beam in a longitudinal
 field configuration with its initial polarization direction
 nominally parallel to the beam axis.
 The general purpose $\mu^+$SR spectrometers at D1 and S1 areas of
 J-PARC with a conventional $^4$He flow cryostat and the Low Temperature
 Facility $\mu^+$SR instrument with a $^3$He-$^4$He dilution
 refrigerator at PSI were used for the temperature ranges of 3-300~K and
 0.1-5~K, respectively. The KTaO$_3$ single crystals were mounted on
 a silver sample holder with the (001) plane perpendicular to
 the beam axis.
 The asymmetry of $\mu^+$ beta decay, $A$, was monitored by the ``Forward''
 and ``Backward'' positron counters as a function of the 
 elapsed time $t$ from the instant of muon implantation.

 $A(t)$ can be decomposed into partial asymmetries from muons stopped in
 the sample, $A_s(t)$, and the silver sample holder, $A_{BG}$, the
 latter of which is substantially a time-independent constant. The
 magnitude of $A_{BG}$ was determined using a reference sample (holmium)
 with similar geometry to the KTaO$_3$ sample and it was
  subtracted from $A(t)$. The remaining $A_s(t)$ was divided by
 $A_s(t=0)$ to obtain the time-evolution of spin polarization $P(t)$ for muons
 that stopped in the sample. In this normalized form, $\mu^+$SR
 data sets recorded with different spectrometers can be directly compared.

 The muons implanted into KTaO$_3$ depolarize mainly due to 
 dipole-dipole interactions with surrounding $^{181}$Ta, $^{39}$K, and
 $^{41}$K nuclei having natural abundances of 99.988\%, 93.258\%, and 6.730\%,
 respectively.\cite{berglund11} $P(t)$ in such a dense nuclear spin
 system can be approximately described by averaging spin precession
 signals for muons that feel random nuclear dipolar fields with an isotropic
 Gaussian distribution (local field approximation).\cite{hayano79}
 When muons are immobile in the $\mu^+$SR time window of up to
 $\sim20~\mu$s, $P(t)$ in a zero
   applied field (ZF) is expected to have a Gaussian-like shape, which
  reflects the Gaussian distribution of the local fields.
 As the hopping rate of muons increases, the shape of $P(t)$ in
 ZF should gradually change from the Gaussian-like curve to an
 exponential-like curve with a decrease in a muon spin relaxation rate 
 (motional narrowing).
  This behavior can be approximately described with 
 the dynamical Gaussian Kubo-Toyabe (DKT) theory based on a strong
 collision approximation.\cite{hayano79} In a faster diffusion regime, the
 effects of muon trapping at dilute defects and/or detrapping from
 them may appear in $P(t)$, which can be approximately expressed using
 a two-state (2S) model.\cite{hempelmann98,ito17,lord01,borghini78}
 
 \section{EXPERIMENTAL RESULTS}\label{secExpResult}

\begin{figure}
\includegraphics[scale =0.51]{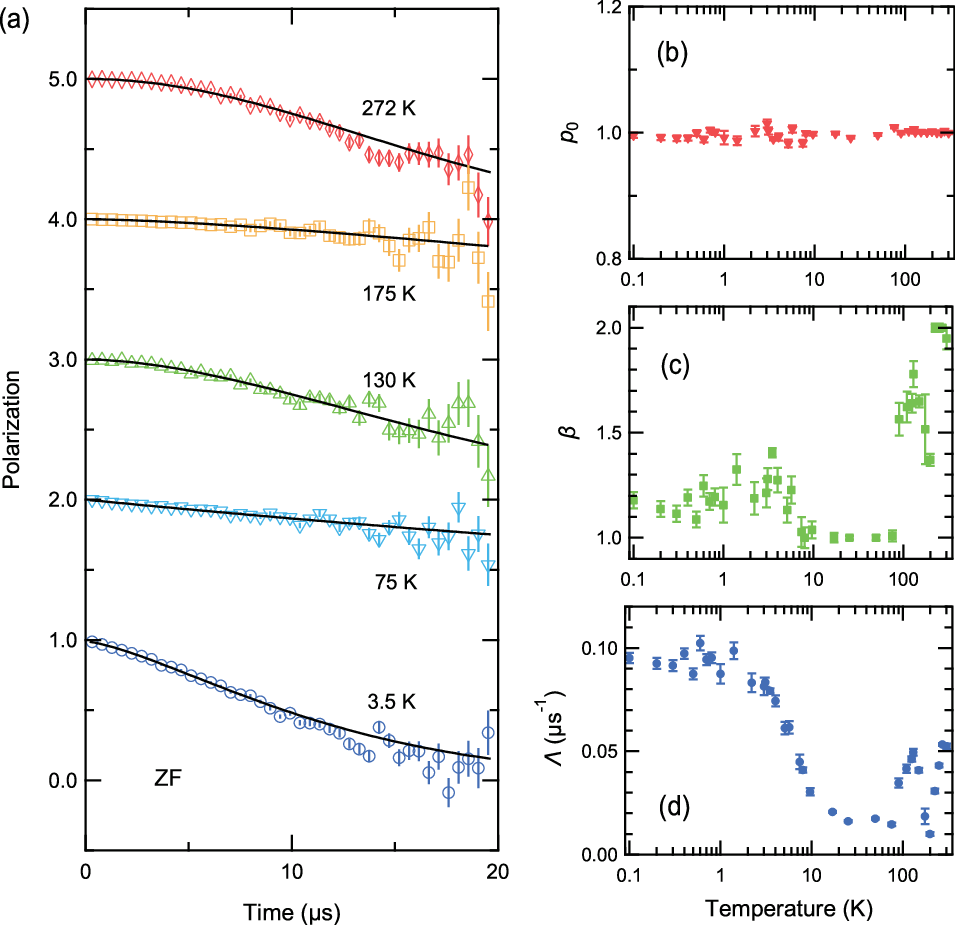}
  \caption{\label{fig1} (a) ZF-$\mu^+$SR spectra at 3.5, 75, 130, 175,
  and 272~K, compared to the best fits to the stretched exponential function, 
  $p_0\exp(-(\Lambda t)^{\beta})$ (solid curves). For clarity, the base
  lines are vertically offset by +1 as the temperature increases.
  Right panels show (b) $p_0$, (c) $\beta$, and (d) $\Lambda$ as
  functions of temperature.}
 \end{figure}

$\mu^+$SR spectra in ZF exhibit a complicated temperature dependence below
room temperature, with the spectral shape changing between Gaussian-like
and exponential-like functions (Fig.\ref{fig1}(a)).
To obtain an overview of the temperature variation, the spectra were
tentatively fitted to a stretched exponential function,
$p_0\exp(-(\Lambda t)^{\beta})$. The initial polarization $p_0$ is
approximately unity over the entire temperature range
(Fig.\ref{fig1}(b)), indicating that almost all the muons implanted into
the sample are in diamagnetic environments. The exponent $\beta$, which
was allowed to vary between 1 and 2, gradually decreases to 1 as the
temperature increases from 1 to 10~K (Fig.~\ref{fig1}(c)). This is
accompanied by a decrease in the muon spin relaxation rate $\Lambda$
(Fig.~\ref{fig1}(d)). This behavior strongly suggests that the hopping
motion of $\mu^+$, which is detectable within the $\mu^+$SR time
window, is activated even below 10~K.
Further information on the $\mu^+$ state in this temperature
regime can be obtained through curve fitting analysis using a DKT
function,\cite{hayano79} which is described in Sec.~\ref{secDKT}.

The exponential-like $P(t)$ at around 10-75 K changes to a Gaussian-like
curve toward 135~K with an increase in $\Lambda$. Similar
behavior has been reported in Sc-doped SrZrO$_3$,\cite{hempelmann98}
BaTiO$_{3-x}$H$_x$ oxyhydride,\cite{ito17} ReO$_3$,\cite{lord01} and
impurity-doped Nb,\cite{borghini78} ascribed to an increasing rate of
$\mu^+$ trapping at a dilute defect with increasing temperature.
Therefore, this behavior in KTaO$_3$ can also be regarded as a signature of
$\mu^+$ trapping at a dilute defect (defect A) that occurs via 
long-range interstitial $\mu^+$ diffusion.
The exponential-like shape is recovered at around 200~K with a
local minimum in $\Lambda$, and then the Gaussian-like shape is restored
with increasing temperature toward 300~K.
The former and latter behavior can be attributed to an increasing rate
of $\mu^+$ jumps between defect A through the interstitial diffusion
path, and that of $\mu^+$ trapping at another more dilute defect
(defect B), respectively.
All these processes above 75~K can be approximately expressed using the 2S
model, which is described in Sec.~\ref{sec2S}.

\subsection{Dynamical Gaussian Kubo-Toyabe analysis for 3.5~$<T<$~25~K}\label{secDKT}

 \begin{figure}
 \includegraphics[scale =0.65]{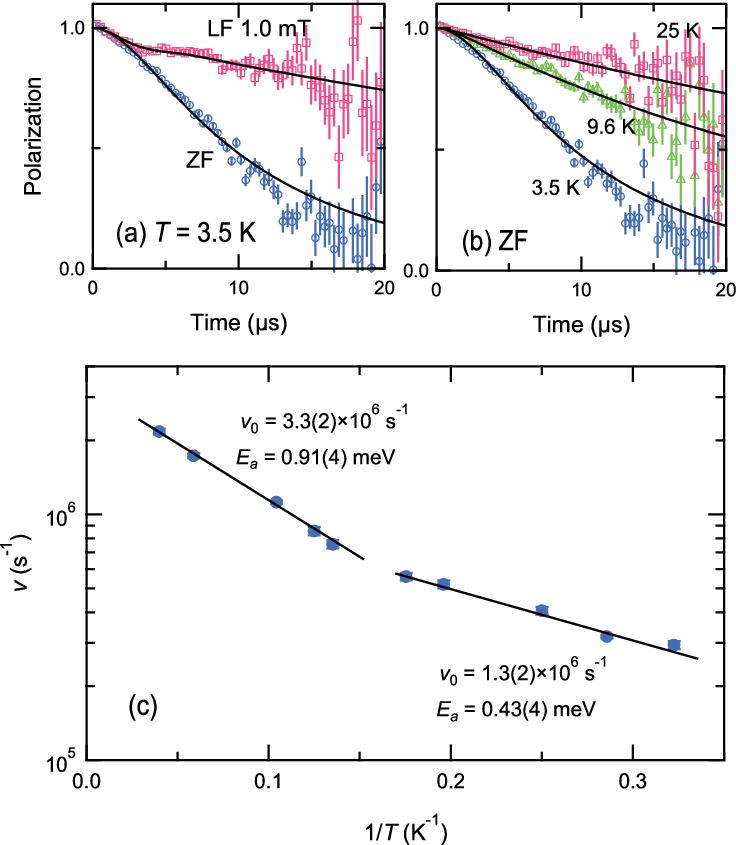}
  \caption{\label{fig2} (a) $\mu^+$SR spectra at 3.5~K under a ZF and a
  longitudinal field of 1.0~mT, compared to the result of the
  simultaneous fit to the DKT function. (b) ZF-$\mu^+$SR spectra at 3.5,
  9.6, and 25~K. Solid curves represent the best fits to the DKT
  function with $\Delta$ fixed to 
  0.1322 $\mu$s$^{-1}$. (c) Arrhenius plot of $\nu$. Solid lines
  represent fits to the Arrhenius-type function.}
 \end{figure}

 \begin{figure}
 \includegraphics[scale =0.6]{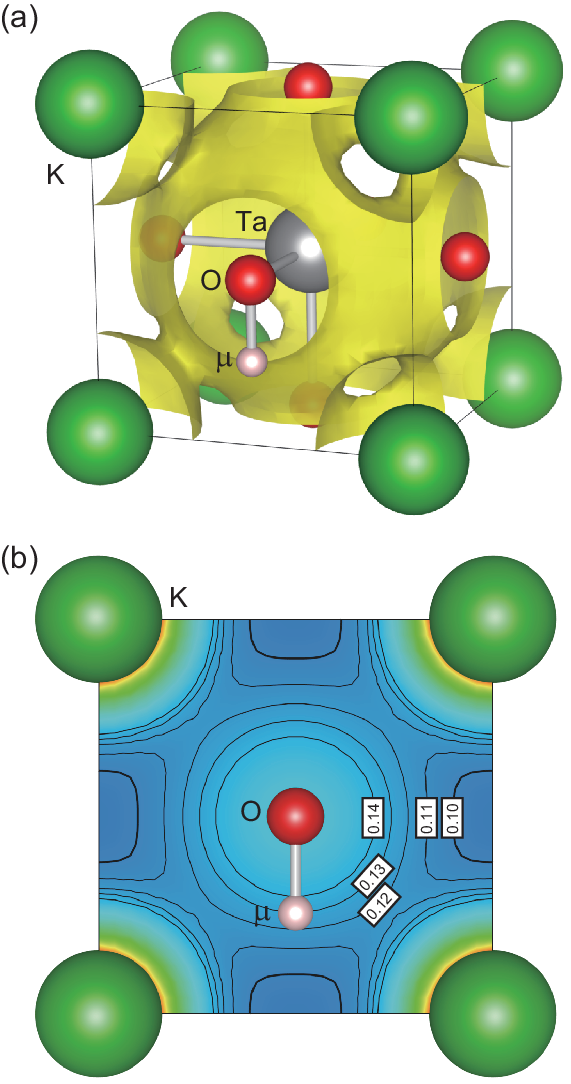}
  \caption{\label{fig3} Spatial distribution of $\Delta_{calc}$ in 
  the KTaO$_3$ lattice. (a) Isosurface for
  $\Delta_{calc}=0.13~\mu$s$^{-1}$. (b) Contour plot
  on the face of the cubic cell. The smallest spheres represent the
  most probable equilibrium position for $\mu^+$ suggested from DFT
  calculations for a hydrogen interstitial impurity in
  KTaO$_3$.\cite{sholl14,gomez07}}
 \end{figure}

The DKT function, $G_{\rm DKT}(t; \Delta,\nu)$, describes
the time evolution of muon spin polarization in fluctuating random local
fields, which are characterized by the root-mean-square (rms) width of
the Gaussian field distribution, $\sigma(=\Delta/\gamma_{\mu})$, where
$\gamma_{\mu}$ is the muon gyromagnetic ratio, and the rate of its
redistribution, $\nu$.
In the present case, $\nu$ can be considered as the $\mu^+$ hopping rate
since the spins of surrounding Ta and K nuclei are substantially static in
the $\mu^+$SR time window. Even when implanted muons induce fast nuclear spin
precession around $\mu^+$-nuclei axes via electric quadrupole
interactions, the nuclear spin components along these axes still produce
the static field distribution at $\mu^+$ sites.

We first performed a simultaneous fit of $\mu^+$SR spectra at 3.5~K under
a ZF and a longitudinal field of 1.0~mT to $G_{\rm DKT}(t; \Delta,\nu)$ and obtained
 $\Delta=0.1322(12)~\mu$s$^{-1}$ (Fig.\ref{fig2}(a)).
The experimental value of $\Delta$ can be compared with a theoretical
value, $\Delta_{calc}$, obtained from a dipolar sum calculation that
takes into account the experimental geometry using the single
crystalline sample.
Figure~\ref{fig3} shows the spatial distribution of $\Delta_{calc}$ in
the KTaO$_3$ lattice. The calculation was performed for the case where
$\mu^+$-induced electric quadrupole interactions on surrounding
nuclei are negligible. The most probable equilibrium position for
 $\mu^+$ suggested by DFT calculations for a hydrogen interstitial
 impurity in KTaO$_3$\cite{sholl14,gomez07} is also shown with the smallest
 sphere, which is very close to the isosurface and contour for
 $\Delta_{calc}=0.13~\mu$s$^{-1}$.
 The values of $\Delta_{calc}$ for $\mu^+$ at K, Ta, and O vacancies
 (V$_{\rm K}$, V$_{\rm Ta}$, and V$_{\rm  O}$) were also calculated for 
 the unrelaxed lattice and
 tabulated in Table~\ref{t1} along with the values of
 $\Delta_{calc}^{eq}$ calculated for the case where the $\mu^+$-induced
 electric quadrupole interactions are taken into account for nearest
 neighbor nuclei. The calculated values for the cation 
 vacancy sites are far from the experimental value, indicating that direct
 $\mu^+$ trapping at these vacancy sites in the $\mu^+$ deceleration
 process can be ignored.
 Based on the good agreement between the experimental and theoretical
 values of $\Delta$ for $\mu^+$ at the interstitial site, we concluded that
 the as-implanted $\mu^+$ mixture is almost purely composed of the
 interstitial $\mu^+$ on the face of the cubic KTaO$_3$ cell, which is
 bound to an oxygen atom to form the O$\mu$ group.
The $\Delta^{eq}_{calc}$ value for the V$_{\rm O}$ site is also close to
the experimental one, but the probability of direct $\mu^+$ trapping to
this site is likely to be negligibly small because H$^-$ is the only
stable hydrogen species at the anion site of perovskite
oxides.\cite{iwazaki10,iwazaki14} To form the muonic analogue of H$^-$,
negatively charged muonium, each muon must acquire two electrons from
the host lattice.\cite{ito20} However, such a process is unlikely to
occur in the final stage of the $\mu^+$ deceleration in the insulating
environment.
  
 Next, ZF-$\mu^+$SR spectra in the temperature range 3-25~K were fitted
 to $G_{\rm DKT}(t; \Delta,\nu)$ with $\Delta$ fixed to 
 0.1322(12)~$\mu$s$^{-1}$ (Fig.~\ref{fig2}(b)),
  which was determined by the simultaneous fit at 3.5~K.
 The temperature variation of $\nu$ is shown in Fig.~\ref{fig2}(c) in
 an Arrhenius form. It appears to obey the Arrhenius law,
 $\nu(T)=\nu_0 e^{-E_a/k_{B}T}$, for two temperature
   ranges of 3.1-5.7~K and 7.4-25.1~K, with $\nu_0=1.3(2)\times 10^6$
 and $3.3(2)\times 10^6$~s$^{-1}$, and $E_a=0.43(4)$ and 0.91(4)~meV,
 respectively, where $E_a$ is the activation energy. The values for the
 preexponential factor $\nu_0$ are much smaller than the Debye frequency of KTaO$_3$, 9.36$\times 10^{12}$~Hz,\cite{bourgeal88}
  suggesting that the interstitial $\mu^+$ diffusion in these temperature
  ranges is not of the classical over-barrier type.
  This may occur through incoherent quantum
 tunneling,\cite{storchak98} but we will not elaborate further on the
 mechanism of the low-temperature motion since quantum tunneling is beyond
 the scope of this paper.

\begin{table}
\begin{ruledtabular}
 \caption{ $\Delta_{calc}$ and $\Delta_{calc}^{eq}$ for $\mu^+$ at
 V$_{\rm K}$, V$_{\rm Ta}$, V$_{\rm  O}$, and the interstitial position
 shown in Fig.~\ref{fig3}.}
% Both $\Delta_{calc}$
% and $\Delta_{calc}^{eq}$ have minima at the midpoint of two K
 % ions, 0.094 and 0.078~$\mu$s$^{-1}$, respectively.}
\label{t1}
\begin{tabular}{ccc}
Muon site~&  ~$\Delta_{calc}$~($\mu$s$^{-1}$)~ & ~$\Delta_{calc}^{eq}$~($\mu$s$^{-1}$)~\\
\hline
 Interstitial & 0.130 & 0.109 \\
 V$_{\rm K}$ & 0.068 & 0.057 \\
 V$_{\rm Ta}$ & 0.045 & 0.044 \\
 V$_{\rm O}$ & 0.172 & 0.142 \\
\end{tabular}
\end{ruledtabular}
\end{table}

\subsection{Two-state model analysis for 75~$<T<$~300~K}\label{sec2S}

\begin{table*}
\begin{ruledtabular}
 \caption{Parameters used for or obtained from the 2S model analysis.}
\label{t2}
\begin{tabular}{cccccc}
Temperature range (K)~&  ~$\Delta_{1}$~($\mu$s$^{-1}$)~ &
	 ~$\Delta_{2}$~($\mu$s$^{-1}$)~& ~$E_a$~(eV)~&
		 ~$\nu_{0}$~(s$^{-1}$)~& ~$c$~\\
\hline
 75-130 & 0.1322 (fixed) & 0.0553(3) & 0.099(4) &
		 1.0(6)$\times$10$^{13}$ & 0.027(4)\\
 150-300 & 0.0553 (fixed) & 0.0583(8) & 0.216(8) &
		 1.3(7)$\times$10$^{12}$ & 0.007(1)\\
\end{tabular}
\end{ruledtabular}
\end{table*}

We adopted a formulation of the 2S model by Lord {\it et al.} used for
describing dynamical $\mu^+$ behavior in ReO$_3$.\cite{lord01}
It assumes that $\mu^+$ starts at $t=0$ in site 1 with a static
relaxation rate $\Delta_1$ and jumps at a rate $\nu$ between equivalent
sites with an activation energy $E_a$. Then, $\mu^+$ finds a trap (site
2) with a static relaxation rate $\Delta_2$ at a probability $c$ per
jump, where the trapped $\mu^+$ stays still for the rest of its lifetime.
Given a static relaxation function $g_1(t)=e^{-\Delta_1^2t^2}$ for 
site 1, the relaxation function $G_1(t)$
for $\mu^+$ that jumps between site 1 and its equivalents and has
never been trapped at site 2 is expressed as,
 \begin{eqnarray}
  &&G_1(t;\nu,c,\Delta_1)  = e^{-\nu(1-c)t}g_1(t;\Delta_1) \nonumber \\ 
  &&+ \nu(1-c)\int_{0}^{t}G_1(t-\tau)e^{-\nu(1-c)\tau}g_1(\tau;\Delta_1)d\tau.\nonumber
 \end{eqnarray}
Using this together with a static relaxation function
$g_2(t)=e^{-\Delta_2^2t^2}$ for site 2, the total relaxation
function $G_{2S}(t)$ including the contribution from $\mu^+$ trapped at
site 2 is given as,
\begin{eqnarray}
 &&G_{2S}(t;\nu,c,\Delta_1,\Delta_2)  = e^{-\nu c t}G_1(t;\nu,c,\Delta_1) \nonumber \\ 
 &&+\nu c\int_{0}^{t}G_1(\tau;\nu,c,\Delta_1)e^{-\nu
  c\tau}g_2(t-\tau;\Delta_2)d\tau. \label{eqn_2S}
\end{eqnarray}

\begin{figure}
 \includegraphics[scale =0.8]{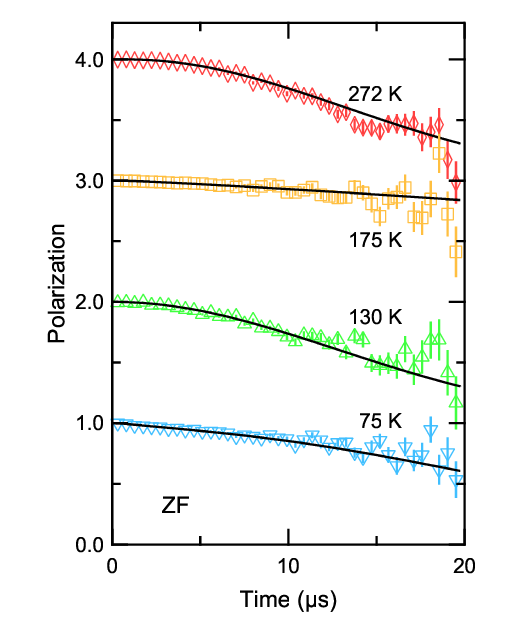}
  \caption{\label{fig4} ZF-$\mu^+$SR spectra at 75, 130, 175, and 272~K,
   compared to the result of the simultaneous fit of the two-state
   model (solid curves). For clarity, the base lines are vertically
   offset by +1 as the temperature increases.}
   \end{figure}

We first performed a simultaneous fit of Eq.~\ref{eqn_2S} to ZF-$\mu^+$SR
spectra in the temperature range 75-130~K, assuming that the jump rate
is subject to the Arrhenius law, $\nu(T)=\nu_0 e^{-E_a/k_{B}T}$.
The result of the fit are shown in Fig.~\ref{fig4} with solid curves.
$\Delta_1$ was fixed to 0.1322~$\mu$s$^{-1}$ through the fit since in
this case site 1 should be associated with the interstitial
bonding site for consistency with the DKT analysis.
$E_a$ and $\nu_0$ for the interstitial $\mu^+$ diffusion, $c$ for
finding defect A, and $\Delta_2$ for $\mu^+$ trapped at defect A were
obtained from the fit, as shown in Table~\ref{t2}.
The value for $E_a$ is close to that for interstitial $\mu^+$ diffusion
in other perovskites, such as SrZrO$_3$\cite{hempelmann98} and
BaTiO$_{3-x}$H$_x$.\cite{ito17}
$\nu_0$ for this temperature range is comparable to the Debye
frequency in contrast to that for the low temperature regime below 25~K.
This suggests that a change in the $\mu^+$ hopping mechanism occurs from the
possible quantum-tunneling type to a classical over-barrier type at
around 75~K with increasing temperature.
The $\Delta_2$ value for $\mu^+$ trapped in defect A is relatively close
to the dipolar sum values for $\mu^+$ at the cation vacancies in
Table~\ref{t1}, suggesting that defect A may be either V$_{\rm K}$ or
V$_{\rm Ta}$.

Equation~\ref{eqn_2S} was also adopted to simultaneously fit
ZF-$\mu^+$SR spectra in the temperature range 150-300~K, where the spectral
shape is expected to primarily reflect $\mu^+$ jumps between defect A
 and trapping in defect B.
Here we assumed that the time that $\mu^+$ spends at the interstitial sites
during its diffusion from one trap to another is negligible in
comparison with that spent in trapped states.
In this analysis, site 1 is supposed to be defect A with
its static relaxation rate of 0.0553~$\mu$s$^{-1}$, which was determined
from the 2S model analysis covering the temperature range 75-130~K.
Therefore, we fixed $\Delta_1$ to this value through the fit.
The result of the fit is shown in Fig.~\ref{fig4} with solid curves.
$E_a$ and $\nu_0$ for the intertrap $\mu^+$ diffusion, $c$ for
finding defect B per intertrap jump, and $\Delta_2$ for defect B were
obtained from the fit, as shown in Table~\ref{t2}.
The value for $E_a$, which means the activation energy for escaping from
defect A in this case, is significantly larger than that
for the interstitial $\mu^+$ diffusion as expected.
$\nu_0$ for this temperature range is close to the Debye
frequency, suggesting that the trapping/detrapping process observed in
this temperature range also occurs through the classical over-barrier hopping
mechanism.
The $\Delta_2$ value for $\mu^+$ trapped in defect B is also close
to the dipolar sum values for $\mu^+$ at the two cation vacancies in
Table~\ref{t1}, suggesting that defect B may be the other one 
of V$_{\rm K}$ and V$_{\rm Ta}$ that was not assigned to defect A.

The primary objective of this paper is to provide a semi-quantitative
description of the long-range interstitial $\mu^+$ diffusion above 75 K 
characterized by $E_a=0.099(4)$~eV and $\nu_0=1.0(6)\times10^{13}$~Hz,
which suggests that it occurs through the classical over-barrier type
hopping mechanism.
In the following sections, we describe DFT calculations performed for
this purpose.

\section{COMPUTATIONAL DETAILS}\label{secComp}
All calculations were performed within the DFT framework using the
Quantum ESPRESSO package.\cite{giannozzi09,giannozzi17} The generalized
gradient approximation using Perdew-Burke-Ernzerhof (PBE) exchange
correlation functional was adopted. Projector augmented-wave potentials
with H($1s$), K(3$s$, 3$p$, 4$s$), Ta(5$s$, 5$p$, 5$d$, 
6$s$) and O(2$s$, 2$p$) valence states were used. Electron wave
functions were expanded in plane waves with a cutoff of 70 Ry for
kinetic energy and 600 Ry for charge density.
The KTaO$_3$ structure was described using a $3\times3\times3$ supercell
of the 5-atom cubic primitive cell.
A H atom was placed at an interstitial position in the host supercell to
model a dilute muon defect in the KTaO$_3$ lattice.
While a 2$\times$2$\times$2 supercell was used for similar calculations
in Ref.~\onlinecite{sholl14,gomez07}, we more carefully employed the
3$\times$3$\times$3 supercell to exclude the possibility of
artificially stabilizing antiferro-distortions, which can occur in the
2$\times$2$\times$2 supercell.
Considering the amphoteric nature of the H defect, calculations were
performed for defective supercells with $q=+1$, 0, and -1, where $q$
is the total charge of electrons and ions in the defective supercell.
For the $q=+1$ or -1 cases, a uniform jellium background with an 
opposite charge was used to maintain charge neutrality.

{\it Structure optimization.} Brillouin zone sampling with a
Monkhorst-Pack {\bf \textit{k}}-point mesh of $2\times2\times2$ 
and Gaussian smearing with a broadening of 0.01 Ry
were applied for structure optimization calculations.
 Atomic positions were relaxed while the lattice constant for the
 supercell was kept fixed at the calculated value of 12.024\AA~ for
 stoichiometric cubic KTaO$_3$, which was determined via full structure
 optimization. This value is consistent with the experimental lattice
 constant of 3.988\AA~ for the 5-atom primitive cell.\cite{samara73} In
 all cases, atomic forces were converged to within $7.7\times10^{-3}$ eV/\AA.

{\it Defect formation energy.} The self-consistent field
calculation for the optimized structure was refined using the optimized
tetrahedron method\cite{kawamura14} with a finer {\bf \textit{k}}-point mesh of
$5\times5\times5$ to avoid artifacts associated with the
Gaussian smearing. The classical formation energy of the interstitial H
defect with a charge $q$, $E_f({\rm H}^q)$, was calculated in accordance with
a standard procedure described in Ref.~\onlinecite{iwazaki10,iwazaki14}:
\begin{eqnarray}
E_f({\rm H}^q) &=& E_t({\rm H}^q) - E_t({\rm host}^0) \nonumber \\ 
 &-& \frac{1}{2}\mu_{{\rm H}_2} + q(\mu_{\rm VBM}+E_{\rm F}) + E_{corr}({\rm H}^q), \nonumber
\end{eqnarray}
where $E_t({\rm H}^q)$ is the total energy of the defective supercell with a
charge $q$, $E_t({\rm host}^0)$ is the total energy of a neutral host
supercell, $\mu_{{\rm H}_2}$ is the chemical potential of an ${\rm H}_2$
molecule, $\mu_{\rm VBM}$ is the chemical potential of the electron
reservoir at the valence-band maximum, and $E_{\rm F}$ is the Fermi
energy with reference to $\mu_{\rm VBM}$.
$E_{corr}({\rm H}^q)$ is a small potential-alignment term for charged defective
supercells,\cite{persson05} which was determined from the difference in
electrostatic potentials at an interstitial position about 10\AA~away
from the H defect.

{\it Classical minimum energy paths.} For the most stable H
 configuration determined by the above calculations, classical minimum
 energy paths (MEPs) between equivalent nearest neighbor H sites were
 searched using the nudged elastic band method\cite{sholl09}
 with seven images. Brillouin zone sampling with a Monkhorst-Pack {\bf
 \textit{k}}-point mesh of $2\times2\times2$ and Gaussian smearing with
 a broadening of 0.01 Ry were applied. Atomic positions were relaxed with
 keeping the lattice constant fixed at the optimized value for
 stoichiometric KTaO$_3$ through iterations for finding the MEPs. A
 climbing image algorithm\cite{henkelman00} was used to identify saddle
 points in the MEPs, which represent transition states (TSs) of
 corresponding H transfer events, and precisely compute classical
 barrier heights $E_0$ at the saddle points.
  In all cases, the norm of the force orthogonal to the path was
 converged to within 0.1 eV/\AA.
    
{\it ZPE-corrected $\mu^+$ jump rates.} The interstitial diffusion of H 
(pseudo) isotopes in the classical over-barrier hopping regime can be
approximately explained using the harmonic transition state theory,
which takes into account isotopic effects arising from 
zero-point vibrations within the harmonic approximation.\cite{sholl14}
We adopted the double Born-Oppenheimer approximation for computing
muon vibrational frequencies, which requires,
in addition to the usual Born-Oppenheimer approximation being satisfied,
that the vibrational motion of muons can be decoupled
from that of other nuclei.\cite{onuorah19}
This is justified because the muon's mass is much lighter than the mass
of any other nucleus.
In the double Born-Oppenheimer and harmonic approximations, ZPEs for muons in the initial
(and final) equilibrium state (ES) and TS are expressed as
$\sum_{i=1}^{3}\frac{1}{2}\hbar\omega_{i}^{\rm ES}$ and
$\sum_{i=1}^{2}\frac{1}{2}\hbar\omega_{i}^{\rm TS}$, respectively, where
$\omega_{i}^{\rm ES}$ and $\omega_{i}^{\rm TS}$ are harmonic vibrational
angular frequencies for muons in the ES and TS.
It should be noted that the sum for the TS runs over only two modes
because the other mode is associated with a
displacement along the MEP and therefore has an imaginary frequency
at the saddle point. 
Consequently, a ZPE-corrected barrier height for the MEP with the 
classical barrier height $E_0$ is given as, 
\begin{equation}
E_{\rm ZPC}  = E_0 + \sum_{i=1}^{2}\frac{1}{2}\hbar\omega_{i}^{\rm TS} - \sum_{i=1}^{3}\frac{1}{2}\hbar\omega_{i}^{\rm ES}. 
\end{equation}
According to the harmonic transition state theory, a ZPE-corrected muon
jump rate $k_{\rm ZPC}$ can be calculated via the following equation,
\begin{equation}
k_{\rm ZPC}  = \frac{1}{2\pi}\frac{\prod_{i=1}^{3}\omega_i^{\rm ES}f(\hbar\omega_i^{\rm ES}/2k_{B}T)}{\prod_{i=1}^{2}\omega_i^{\rm TS}f(\hbar\omega_i^{\rm
 TS}/2k_{B}T)}\exp(-E_0/k_{B}T), \label{eqn_hTST}
\end{equation}
where $f(x)=\sinh(x)/x$.\cite{sholl14}
At high temperatures, $f(\hbar\omega/2k_{B}T)\rightarrow 1$, so that
eq.~\ref{eqn_hTST} reduces to the simple Arrhenius law with the
activation energy of $E_0$. 
At low temperatures,
\begin{equation}
k_{\rm ZPC} \rightarrow  (k_{B}T/h)\exp(-E_{\rm ZPC}/k_{B}T), 
\end{equation}
where the preexponential factor $k_{B}T/h$ is in the order of
10$^{12}$~Hz in the temperature range 75-300~K, which is close to the
Debye frequency of KTaO$_3$.
To calculate $k_{\rm ZPC}$ via eq.~\ref{eqn_hTST}, $\omega_{i}$s for the ES
and TSs in all possible MEPs were computed using a density functional
perturbation theory module in the Quantum ESPRESSO package within the
framework of the double Born-Oppenheimer and harmonic approximations. In the density functional
perturbation theory calculations, the muon was treated as a H isotope
having the mass of 0.113$m_p$, where $m_p$ is the proton's mass.

\section{COMPUTATIONAL RESULTS}\label{secCompRes}
\subsection{Classical defect formation energy}
\begin{figure}
 \includegraphics[scale =0.8]{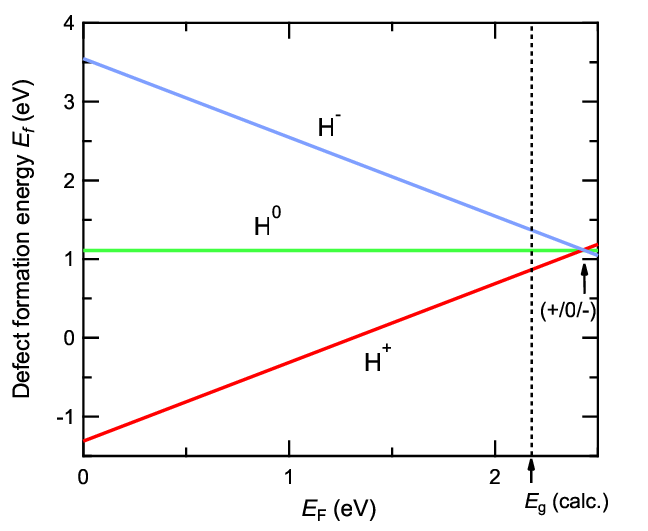}
  \caption{\label{fig5} Classical defect formation energies $E_f({\rm
  H}^q)$ for $q=+1$, 0, and $-1$ as functions of $E_{\rm F}$. $E_g$ is
  the bandgap calculated for a non-defective KTaO$_3$ lattice.}
 \end{figure}

Classical defect formation energies $E_f({\rm H}^q)$ for $q=+1$, 0, and
$-1$ are obtained as functions of $E_{\rm F}$, as shown in Figure~\ref{fig5}.
The three $E_f({\rm H}^q)$ lines cross at an $E(+/0/-)$
transition level in the conduction band. This means that 
the interstitial H always behaves as a shallow donor regardless of the
position of $E_{\rm F}$.
All of the $q=+1$, 0, and $-1$ solutions have the OH group with its
symmetry axis along the $\langle 001\rangle$ direction, which is
consistent with the previous DFT studies.\cite{sholl14,gomez07}
It should be noted that in the $q=0$ and $-1$ cases, excess electrons
are not tightly bound to H, but are delocalized in the bottom of the
conduction band.
We carefully explored the possibility of electron localization using the
PBE+$U$ method with a Hubbard $U$ parameter of 5 eV for Ta 5$d$ states,
but this did not change the results.
Therefore, all the solutions mean that the
oxygen-bound protonic state is most stable in KTaO$_3$.

The displacements of the K, Ta, and O atoms furthest from the proton at
the equilibrium position in the $q=+1$ condition are as small as
0.008, 0.043, and 0.018\AA, respectively, with reference to
corresponding atomic positions in the non-defective cubic lattice,
confirming that the 3$\times$3$\times$3 supercell is sufficiently large.

\begin{table*}
\begin{ruledtabular}
 \caption{Harmonic vibrational angular frequencies $\omega_i$ 
 %(in cm$^{-1}$ unit, which is conventionally used in optical spectroscopies)
 and corresponding harmonic ZPEs for muons in the ES, TS(T), and TS(R),
 together with the classical barrier heights $E_{\rm 0}$ obtained from
 the nudged elastic band calculations and the ZPE-corrected barrier
 heights $E_{\rm ZPC}$ for the two paths.} 
\label{t3}
\begin{tabular}{ccccccc}
Muon's state ~&  ~$\omega_{1}$~(cm$^{-1}$)~ &
	 ~$\omega_{2}$~(cm$^{-1}$)~ & ~$\omega_{3}$~(cm$^{-1}$)~ &
		 ZPE~(eV) & $E_{\rm0}$~(eV) & $E_{\rm ZPC}$~(eV) \\
\hline
 ES & 10306 & 2528 & 2288 & 0.9375 & - & - \\
 TS(T) & 5120 & 3844 & 3780$i$ & 0.5558 & 0.4333 & 0.0516 \\
 TS(R) & 11054 & 2500 & 1733$i$ & 0.8403 & 0.2185 & 0.1213 \\
\end{tabular}
\end{ruledtabular}
\end{table*}

 \subsection{Classical minimum energy paths for $\mu^+$ by the nudged elastic band method}

\begin{figure}
 \includegraphics[scale =0.70]{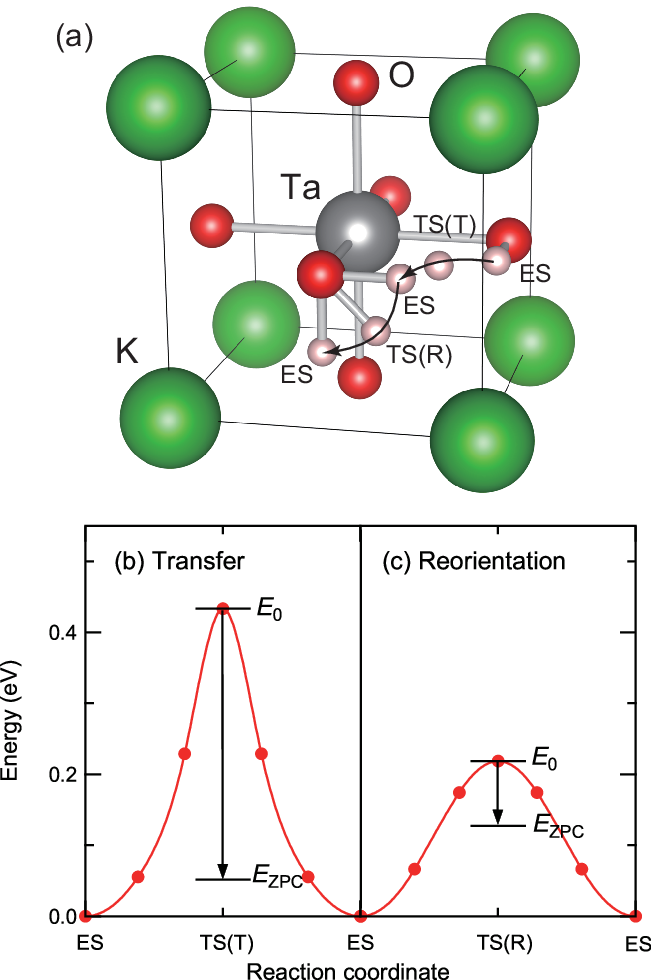}
  \caption{\label{fig6} (a) Schematic illustration of minimum energy
  paths for muon diffusion in KTaO$_3$. (Bottom) One-dimensional energy
  profiles for muons moving along the transfer-type minimum energy path (b) and
  the reorientation-type minimum energy path (c), estimated using the
  nudged elastic band method. The reductions of the effective barrier heights due to the
  harmonic ZPE correction are represented with arrows.}
 \end{figure}

The global proton migration in the KTaO$_3$ lattice can be
decomposed into two elementary steps of H hopping between equivalent ES
sites,\cite{sholl14,gomez07} as shown in Fig.~\ref{fig6}(a). One is the  
H/$\mu$ transfer process between adjacent oxygen atoms via a bond-breaking
transition state TS(T). The other is the H/$\mu$ reorientation process
around an oxygen atom via a transition state TS(R) without breaking the
OH/O$\mu$ bond. One-dimensional energy profiles for H/$\mu$ moving
along the transfer-type and reorientation-type MEPs are shown in
Fig.~\ref{fig6}(b) and (c), respectively. 
The bottleneck process of the long-range interstitial diffusion is the 
 H/$\mu$ transfer ($E_0=0.43$~eV).
The activation barrier height without the ZPE correction is
approximately four times higher than the experimental value of $\sim 0.1$~eV.

 \subsection{Zero-point energy corrected $\mu^+$ jump rates}
The parameters for the harmonic ZPE correction, $\omega_{i}$ and
corresponding ZPEs for the ES, TS(T), and TS(R), are shown in Table~\ref{t3}.
A marked difference between ZPEs for the ES and TS(T) reflects the fact that
the strong O$\mu$ bond associated with especially large $\omega_i$s is temporarily
broken in the TS(T). Consequently, the reduction in the effective barrier
height due to the ZPE correction is more pronounced in the
transfer-type path, resulting in a reversal of $E_{\rm ZPC}$ between the two
paths [Table~\ref{t3} and Fig.~\ref{fig6}(b,c)].
The ZPE correction based on the harmonic transition state theory leads
to the bottleneck barrier height of $E_{\rm ZPC}=0.1213$~eV for the
reorientation-type path, which is close
to the experimental value of $\sim 0.1$~eV for long-range interstitial
$\mu^+$ diffusion in KTaO$_3$, as well as Sc-doped
SrZrO$_3$\cite{hempelmann98} and BaTiO$_{3-x}$H$_x$.\cite{ito17}

\begin{figure}
 \includegraphics[scale =0.7]{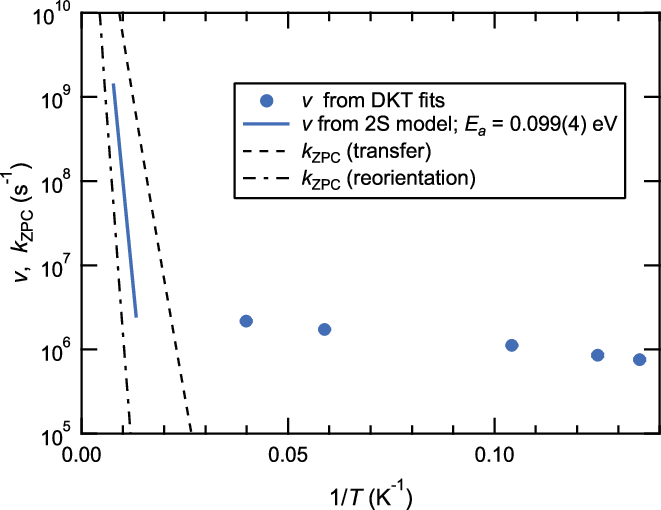}
  \caption{\label{fig7} Arrhenius plot for $\nu$ from the experiment and
$k_{\rm ZPC}$ calculated with Eq.~\ref{eqn_hTST}.}
 \end{figure}

A comparison between $\nu$ from the experiment and
$k_{\rm ZPC}$ calculated with Eq.~\ref{eqn_hTST} is made in Fig.~\ref{fig7}.
The variation in the slope of $\nu(1/T)$ suggests that a change in
the $\mu^+$ hopping mechanism occurs from the possible quantum-tunneling
type to the classical over-barrier type at around 75~K with increasing
temperature. $k_{\rm ZPC}$ for the bottleneck path (dash-dotted line)
approximately reproduces the temperature dependence of $\nu$ in the
over-barrier hopping regime.

\section{DISCUSSION}\label{secDisc}
KTaO$_3$ exhibits quantum paraelectric behavior at low temperatures,
where the cubic structure and a virtual ferroelectric structure are
nearly degenerate and the cubic symmetry is maintained via zero-point
quantum fluctuations over these configurations. The atomic displacement
in the virtual ferroelectric phase from corresponding atomic positions
in the cubic phase was estimated from DFT calculations to be at most
0.026\AA.\cite{esswein22} This is one order of magnitude smaller than
the local displacement of muon-neighboring atoms from our calculations,
which reflects the chemical bonding state with muons. Therefore, the
dynamical ferroelectric displacement associated with the quantum
paraelectricity is likely to have only minor effects on muon kinetics,
which involves the formation and breaking of the chemical bonds.

Since the potentials experienced by oxygen-bound protons in perovskite
oxides generally reveal significant anharmonic terms along some
directions,\cite{iwazaki10} it is not obvious whether the harmonic
method can serve as a good approximation for describing ZPEs of muons in
KTaO$_3$. In particular, the adiabatic potentials corresponding to the
O$\mu$ stretching vibrations (the $\omega_1$ modes for ES and TS(R) in
Table~\ref{t3}) are highly asymmetric and anharmonic, and the deviation
from the quadratic curve for the harmonic approximation should be
significant except at the bottom of the potential wells.

A similar situation has been reported for vibrational modes of muons at
around equilibrium positions in some fluorides.\cite{moller13-2} In that
paper, a comparison was made between harmonic ZPEs and those obtained by
solving the Schr\"{o}dinger equation for a muon in one-dimensional adiabatic
potentials calculated along each vibration direction. Despite
significant anharmonicity identified in some vibrational modes, there
was surprisingly good agreement between the two sets of ZPEs.

We adopted the same strategy for KTaO$_3$ and solved the Schr\"{o}dinger
equation for a muon in adiabatic potentials corresponding to the seven
muon vibrational modes in Table \ref{t3} with real frequencies using a
finite-differences method. As a result, we obtained ZPEs of 0.9632,
0.5488 and 0.8621 eV for ES, TS(T), and TS(R), respectively, and the
bottleneck barrier height of 0.1173 eV. The good agreement with
corresponding harmonic values in Table \ref{t3} strongly suggests the
validity of the harmonic approximation for describing the vibrational
states of muons in perovskite oxides.

The ZPE correction based on the harmonic 
approximation successfully accounted, at least semi-quantitatively, for
the long-range interstitial $\mu^+$ diffusion in KTaO$_3$ with the
activation energy of $\sim 0.1$~eV.
This relatively low-cost approach is probably also applicable to
explaining that in Sc-doped SrZrO$_3$ and BaTiO$_{3-x}$H$_x$, which is 
characterized by similar activation energies.\cite{hempelmann98,ito17}
Furthermore, we can gain general insight into $\mu^+$ states in oxides
from the calculation results for KTaO$_3$ as described below.

The long-range interstitial $\mu^+$ diffusion in
oxides can usually be expressed as the combination of the transfer and
reorientation-type hopping processes as in the case of KTaO$_3$.
The transfer-type process is characterized by the temporary breaking of
the strong O$\mu$ bond in the TS(T), which results in a marked difference
between ZPEs for the ES and TS(T). This can not only cause a significant 
reduction in the effective barrier height, but also change the ES for
muons in extreme cases; from a viewpoint of ZPEs, muons prefer to stay in
large interstitial or vacancy sites rather than forming the strong
O$\mu$ bond at the conventional proton site adjacent to oxygen.
This means that muons and hydrogen may be found in different stable
sites due to the ZPE effect.

The barrier height for the transfer-type path between the conventional
proton sites can vary widely depending on host crystal structures. When
the proton sites are dense in the host lattice, as in the KTaO$_3$
perovskite, the barrier height 
can be relatively low in comparison with dilute cases; when the proton
sites are far apart from each other, a higher energy for {\it completely}
breaking the O$\mu$ bond ($\sim$5~eV, estimated from the average OH bond
energy in H$_2$O) is expected to be required for
transferring a muon from one site to another.
Therefore, whether {\it global} muon diffusion becomes important in 
muon spin relaxation below room temperature should strongly depend on
the crystal structure of host oxides.

On the other hand, the energy required for transferring a muon through
the reorientation-type path is considered to be relatively independent
of the crystal structure of host oxides because this
process does not involve the temporary breaking of the O$\mu$ bond.
Thus, provided that the crystal symmetry allows for such reorientation,
the local motion of muons around oxygen may be characterized by a small
barrier height of 0.1$\sim$0.2~eV. Since $\mu^+$SR is a local
probe, even such local motion can cause significant changes in $\mu^+$SR 
spectra.\cite{dehn20,dehn21} Therefore, researchers who use $\mu^+$SR to study
oxides need always keep in mind the possibility that muons can move even
below room temperature when examining $\mu^+$SR data, and then carefully
discuss physical properties of the oxides, such as magnetism and ionic
conductivity.

It is worth mentioning that the above discussion on global and local
motion of muons in oxides may also be applied to materials that involve
elements with high electronegativity, such as C, N, S, Se, and halogens,
which tend to create strong chemical bonds with implanted
muons.\cite{brewer86,nishiyama03,lancaster07}
It should also be noted that quantum tunneling of muons
is beyond the scope of the DFT and harmonic transition state theory
framework and higher-level calculations are necessary for correctly
handling this phenomenon, such as using the {\it ab initio}
path-integral molecular dynamics method.\cite{kimizuka18}

The harmonic ZPEs for muons can be converted to those for protons under
the double Born-Oppenheimer approximation by multiplying by the square root of the
muon-proton mass ratio ($\sqrt{m_{\mu}/m_{p}}\sim$ 1/3). From these
scaled values, the bottleneck process for long-range diffusion of
protons is inferred to remain the transfer type, with an activation
barrier of about 0.3~eV. This value is considerably larger than that for
muons, suggesting that the over-barrier-type diffusion of hydrogen
(pseudo) isotopes in KTaO$_3$ exhibits a normal kinetic isotope
effect. However, the justification for applying the double Born-Oppenheimer approximation to
protons remains to be established, and more careful treatment may be
necessary.

\section{CONCLUSIONS}\label{secConc}

In summary, we carried out $\mu^+$SR and DFT studies on $\mu^+$
diffusion in the cubic perovskite KTaO$_3$ to gain a comprehensive
understanding of the $\mu^+$ diffusion in the perovskite oxide lattice.
In the $\mu^+$SR experiment, we observed motional narrowing
behavior, the evidence of interstitial $\mu^+$ diffusion, at cryogenic
temperatures. 
The $\mu^+$ diffusion observed below room temperature was classified
into two types; one is possibly the quantum-tunneling type, which is
dominant below 75~K, and the other is the classical over-barrier type,
which shows a major contribution at higher temperatures.
The latter was indirectly identified through interactions between muons
and defects, characterized by a small activation energy of $\sim 0.1$~eV.
We successfully accounted, at least semi-quantitatively, for the small
activation energy associated with the long-range interstitial $\mu^+$ 
diffusion using the DFT and harmonic transition state theory 
framework. The significant reduction in the effective barrier 
height for the transfer-type $\mu^+$ hopping path due to the ZPE
correction is the key to understanding the small activation energy for
KTaO$_3$ and other perovskite oxides.

We also obtained important insight into $\mu^+$ states in other oxides
from the calculation results for KTaO$_3$. The effective barrier height
for the transfer-type path can vary widely depending on the crystal
structure of host oxides, whereas that for the reorientation-type path
is expected to be relatively structure-independent and in the range of
approximately 0.1 to 0.2~eV if the crystal symmetry allows for such
reorientation. 
The low barrier height suggests that significant changes in $\mu^+$SR 
spectra can arise even below room temperature due to the local
$\mu^+$ motion regardless of whether the global motion is activated or not.
This may prompt a revision of previous interpretations of
$\mu^+$SR anomalies in some oxides, which have been claimed to
be directly associated with the intrinsic properties of the materials.

\begin{acknowledgments}
We thank the staff of the J-PARC and PSI muon facilities for technical
 assistance and A.~Koda, R.~Kadono, and K.~Fukutani for helpful discussions.
The DFT calculations were conducted with the supercomputer HPE
SGI8600 in Japan Atomic Energy Agency. This research was partially
 supported by Grants-in-Aid (Nos. 23K11707, 21H05102, 20K12484, 20H01864, and
 20H02037) from the Japan Society for the Promotion of Science.
\end{acknowledgments}

\end{document}